**ARTICLE**  Open Access

# A 3.584 Tbps coherent receiver chip on InP-LiNbO$_3$ wafer-level integration platform

Xiaojun Xie[1]✉, Chao Wei[1], Xingchen He[1], Yake Chen[1], Chenghao Wang[1], Jihui Sun[1], Lin Jiang[1], Jia Ye[1], Xihua Zou[1], Wei Pan[1] and Lianshan Yan[1]✉

**Abstract**
The rapid advancement of the thin-film lithium niobate (LiNbO$_3$) platform has established it as a premier choice for high-performance photonics integrated circuits. However, the scalability and cost-efficiency of this platform are hindered by the reliance on chip-level fabrication and integration for passive and active components, necessitating a robust wafer-level LiNbO$_3$ heterogeneous integration platform. Despite its critical role in enabling ultrahigh-speed optical interconnects, as well as optical mmWave/THz sensing and communication, the realization of ultrahigh-speed photodiodes and optical coherent receivers on the LiNbO$_3$ platform remains an unresolved challenge. This is primarily due to the challenges associated with the large-scale integration of direct-bandgap materials. To address these challenges, we have developed a scalable, high-speed InP-LiNbO$_3$ wafer-level heterogeneous integration platform. This platform facilitates the fabrication of ultrahigh-speed photodiodes with a bandwidth of 140 GHz, capable of receiving high-quality 100-Gbaud pulse amplitude modulation (PAM4) signals. Moreover, we demonstrate a seven-channel, single-polarization I–Q coherent receiver chip with an aggregate receiving capacity of 3.584 Tbit s$^{-1}$. This coherent receiver exhibits a balanced detection bandwidth of 60 GHz and a common mode rejection ratio (CMRR) exceeding 20 dB. It achieves receiving capacities of 600 Gbit s$^{-1}$ λ$^{-1}$ with a 100-Gbaud 64-QAM signal and 512 Gbit s$^{-1}$ λ$^{-1}$ with a 128-Gbaud 16-QAM signal. Furthermore, energy consumption as low as 9.6 fJ bit$^{-1}$ and 13.5 fJ bit$^{-1}$ is achieved for 200 Gbit s$^{-1}$ and 400 Gbit s$^{-1}$ capacities, respectively. Our work provides a viable pathway toward enabling Pbps hyperscale data center interconnects, as well as optical mmWave/THz sensing and communication.

## Introduction

The rapid development of cloud-based artificial intelligence (AI) services is driving data traffic in hyperscale data center clusters to the Pbps scale[1]. This surge in data traffic has created a strong demand for ultra-large capacity solutions. High-speed interconnects within a hyperscale data center are essential for optimal system performance. However, traditional electrical interconnects are constrained by physical limitations, which restrict their bandwidth, power efficiency, and latency in hyperscale systems[2]. Integrated photonic interconnects present a promising alternative, offering higher bandwidth, lower latency, and improved energy efficiency[3]. The performance of electrical-optical (E-O) and optical-electrical (O-E) conversions is critical for integrated photonic interconnects. A 300 Gbaud E-O and O-E interface with an analog bandwidth of >150 GHz in coherent transceivers is required to meet the requirements of multiple Tbps and Pbps per lane optical interconnect[4]. In addition to the hyperscale AI clusters, the ongoing advancements in 6G communication research are driving a significant shift toward higher frequency ranges, including the millimeter-wave (mmWave) and terahertz (THz) bands[5]. This transition is driven by the spectrum congestion in lower frequency ranges and the vast contiguous bandwidth available in the mmWave and THz regimes, which is essential to meet the escalating demands for ultra-high data rates and low-latency communication in next-generation networks. Photonics technology offers significant advantages for mmWave and

Correspondence: Xiaojun Xie (xxie@swjtu.edu.cn) or
Lianshan Yan (lsyan@home.swjtu.edu.cn)
[1]Key Laboratory of Photonic-Electronic Integration and Communication-Sensing Convergence, School of Information Science and Technology, Southwest Jiaotong University, 611756 Chengdu, China
These authors contributed equally: Xiaojun Xie, Chao Wei.

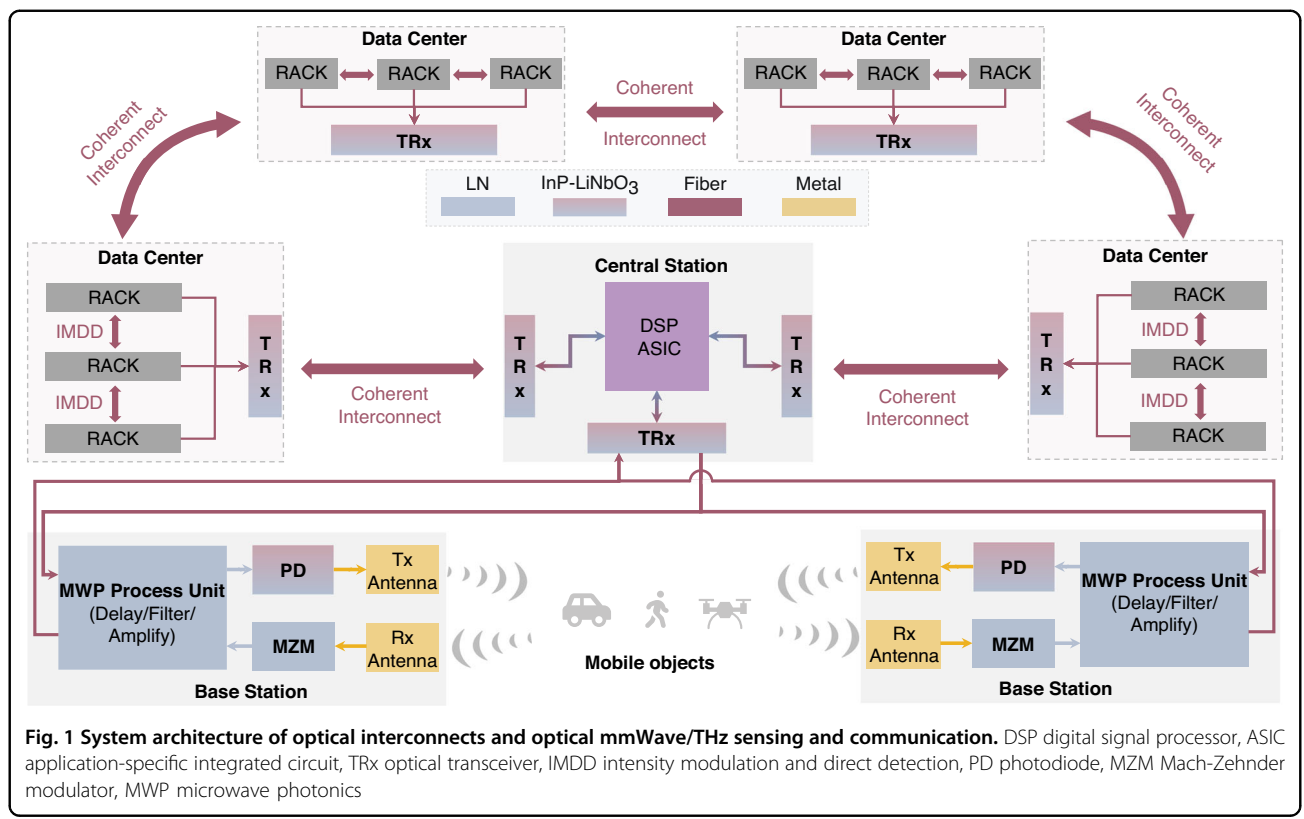

**Fig. 1 System architecture of optical interconnects and optical mmWave/THz sensing and communication.** DSP digital signal processor, ASIC application-specific integrated circuit, TRx optical transceiver, IMDD intensity modulation and direct detection, PD photodiode, MZM Mach-Zehnder modulator, MWP microwave photonics

THz sensing and communication systems, including large modulation bandwidth, low-noise mmWave/THz signal generation, and true time delay capabilities[6]. Figure 1 illustrates the applications and system architecture of optical interconnects and optical mmWave/THz sensing and communication.

However, the typical bandwidths of EO modulation and OE detection on photonics integration platforms, including silicon on insulator (SOI)[7–9], silicon nitride ($Si_3N_4$)[10–12], and indium phosphide (InP)[13–15], are limited to tens of gigahertz. Thin-film $LiNbO_3$ has emerged as a high-performance photonics integration platform, particularly for high-speed applications, due to its strong electro-optic coefficient, tight mode confinement, and wide transparency window[16–19]. The state-of-the-art thin-film $LiNbO_3$ I–Q modulator, benefiting from a large Pockels coefficient of 32 pm·$V^{-1}$, has demonstrated exceptional performance with a bandwidth exceeding 110 GHz and a half-wave voltage ($V_\pi$) of less than 1 V[20]. Theoretically, the bandwidth of a thin-film $LiNbO_3$ modulator is estimated to exceed 300 GHz[21]. Besides, several electrically-pumped lasers[22–24] and high-speed photodiodes[25,26] have been demonstrated on thin-film $LiNbO_3$ platforms using hybrid and heterogeneous techniques[27–32]. Despite the significant advancements in $LiNbO_3$ integrated devices to date, widespread adoption of $LiNbO_3$ integrated photonics remains limited by several key factors. Firstly, the fabrication of these devices typically relies on electron beam lithography, a chip-level process that poses substantial challenges for large-scale, cost-effective production. Secondly, the heterogeneous integration of active devices on thin-film $LiNbO_3$ is also predominantly performed at the chiplet level. Developing a wafer-level heterogeneous integration process for active devices on the $LiNbO_3$ platform is crucial to enable massively scalable thin-film $LiNbO_3$ photonics integrated circuits. Thirdly, as a core component for ultra-large capacity optical interconnects[33], ultra-high speed photonics computing[34,35], and high-performance microwave/mmWave/THz photonics[36,37], ultrahigh-speed photodiodes and coherent receiver chips have not been reported on the thin-film $LiNbO_3$ integration platform due to the absence of a reliable wafer-level heterogeneous integration platform. Furthermore, a critical gap remains in achieving the monolithic integration of mmWave/THz modulators, photodetectors, and coherent receivers on the same chip to meet the stringent performance requirements of mmWave/THz photonic applications.

In this article, to the best of our knowledge, we present the first demonstration of an ultrahigh-speed coherent receiver chip and its single-chip integration with high-speed modulators on a scalable InP-$LiNbO_3$ wafer-level integration platform. A single chip with seven-channel single-polarization I–Q coherent receivers demonstrates an aggregate data capacity of 3.584 Tbit s$^{-1}$. The I–Q coherent receiving chip consists of a compact high-performance 2 × 4 90° optical hybrid and a high-speed balanced PD array. The single channel coherent receiver achieves reception of 128 Gbaud 16-QAM signal (512 Gbit s$^{-1}$) and 100 Gbaud 64-QAM signal (600 Gbit s$^{-1}$) signals. The 100 Gbaud QPSK and 16-QAM signals are demodulated successfully after transmissions of 1040 km and 25 km, respectively. Notably, ultra-low energy consumption of 9.6 fJ bit$^{-1}$ and 13.5 fJ bit$^{-1}$ is achieved for capacities of 200 Gbit s$^{-1}$ and 400 Gbit s$^{-1}$, respectively. The performance of this single-polarization coherent receiver outperforms all previously demonstrated integrated coherent receiver chips, setting new records in bandwidth, capacity, and energy efficiency. Our work provides a potential solution for Pbps hyperscale data center interconnects, microwave/mmWave/THz photonics signal generation and manipulation, and full-photonics mmWave/THz integrated sensing and communication.

## Results
### Fabrication

Figure 2a presents the schematic diagram of the wafer-level heterogeneous integration platform. Single PDs, PD arrays, balanced PDs, optical hybrids, intensity modulators, and coherent receivers were designed on a reticle with a size of 1 × 1 cm. Compared with our previous epitaxial layers[26,38], the epitaxial layers in this work are designed to form an n-mesa-down structure after wafer bonding. This design aims to reduce the series resistance, improve the bandwidth, and increase the output radio frequency power. The simulated series resistances of the heterogeneous integrated photodiode with an n-mesa-down structure and p-mesa-down structures are detailed in Supplementary Note 1. The epitaxial layers of the InP/InGaAs wafer were grown on a semi-insulating InP substrate using metal-organic chemical vapor deposition (MOCVD), as detailed in Supplementary Note 1.

The main fabrication process is depicted in Fig. 2b. Initially, thin-film LiNbO$_3$ waveguides and passive devices were fabricated at the wafer level. Low-loss ridge waveguides and high-performance multimode interference (MMI) couplers were formed using argon-based dry etching. The thin-film LiNbO$_3$ waveguide had a total thickness of 600 nm and a slab thickness of 300 nm. False-colored microscope images of the fabricated waveguide and 1×N MMI couplers are shown in Fig. 2c. Following the dry etching of the passive components, the thin-film LiNbO$_3$ wafer was thoroughly polished and cleaned. Subsequently, a 2-inch InP/InGaAs wafer was bonded onto the thin-film LiNbO$_3$ wafer. Selective wet etching with hydrochloric acid (HCl) was used to remove the InP substrate. The wet etching was precisely stopped at the InGaAs p-contact layer. The metal was then deposited onto the p-contact layer. The PD active region employed a double-mesa structure, with the p mesa etched down to the n-contact layer using a chlorine-based dry etching process. The same dry etching process was used to etch the n mesa. N-contact metal deposition was performed immediately to prevent any possible oxidation. Wet etching was then applied to expose the LiNbO$_3$ surface, preventing potential damage. A 600 nm SiO$_2$ layer was deposited over the entire wafer surface as a passivation layer to reduce the dark current and protect the waveguides and passive devices from impurities in the subsequent processes. The SiO$_2$ above the p- and n-contact metal stacks was dry-etched to form the metal via. Metal electrodes were created by electroplating and lift-off. After PD fabrication, the SiO$_2$ passivation layer in the modulator region was etched to expose the LN waveguide. The metal was deposited on the LN to form a transmission line. A SiO$_2$ cladding layer was deposited on the whole wafer, and then the SiO$_2$ cladding layer on the pad region was opened by dry etching. The wafer was then cut into 1 × 1 cm chips, which were side-polished to reduce coupling loss. More details are provided in the "Methods" section.

### Device design and characteristics

1 × N MMI couplers, operating in self-imaging principle, were used to evenly distribute light in the 1 × N PD array. The dimensions of the 1 × 2, 1 × 3, and 1 × 4 MMI couplers were 6 × 26 μm, 6 × 38 μm, and 6 × 52 μm, respectively. The fabrication yield of the wafer-level heterogeneous integration is estimated to be greater than 80% based on the I–V characterization of more than 200 randomly selected single PDs from the same wafer. Figure 2d shows the statistics of the dark currents at a bias voltage of −4 V. Most of the devices exhibited dark currents below 300 nA, with some having dark currents as low as a few nA (see Supplementary Note 2). Figure 2e illustrates the schematic diagram and performance of the fabricated intensity modulator. Based on the measured S parameters, the characteristic impedance and microwave effective index were extracted as 45 Ω and 2.15, respectively. The intensity modulator exhibited a > 50-GHz 3-dB bandwidth and a 4-V half-wave voltage. More details about the intensity modulator can be found in Supplementary Note 3.

The bandwidth and output RF power was measured using heterodyne techniques, as detailed in Supplementary Note 4. The frequency response of the PDs was characterized at a fixed photocurrent using four RF probes covering the DC–67 GHz, 75–110 GHz, 90–140 GHz, and 110–170 GHz frequency bands. A 1.5 × 8 μm single PD exhibited a 3-dB bandwidth of 140 GHz with 0.4 A W$^{-1}$ responsivity, as shown in Fig. 2f.

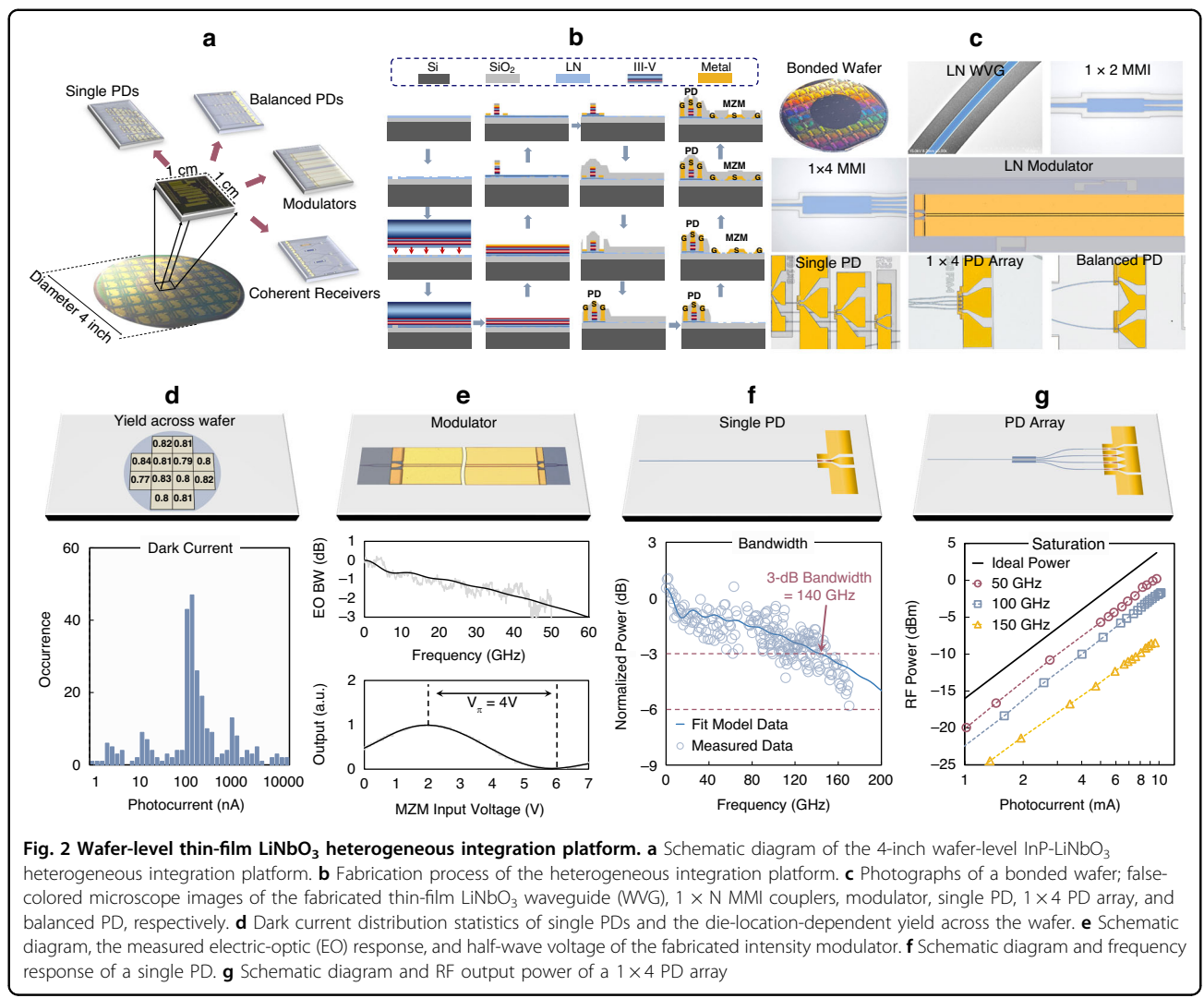

**Fig. 2 Wafer-level thin-film LiNbO$_3$ heterogeneous integration platform. a** Schematic diagram of the 4-inch wafer-level InP-LiNbO$_3$ heterogeneous integration platform. **b** Fabrication process of the heterogeneous integration platform. **c** Photographs of a bonded wafer; false-colored microscope images of the fabricated thin-film LiNbO$_3$ waveguide (WVG), 1 × N MMI couplers, modulator, single PD, 1 × 4 PD array, and balanced PD, respectively. **d** Dark current distribution statistics of single PDs and the die-location-dependent yield across the wafer. **e** Schematic diagram, the measured electric-optic (EO) response, and half-wave voltage of the fabricated intensity modulator. **f** Schematic diagram and frequency response of a single PD. **g** Schematic diagram and RF output power of a 1 × 4 PD array

More details about the responsivity can be found in Supplementary Note 5. The S-parameters of the device were measured using a vector network analyzer with a frequency range of DC–67 GHz. By fitting the measured S-parameters to the equivalent circuit model (see Supplementary Note 6), the extracted capacitance and resistance were 11.5 fF and 12 Ω, respectively. The parasitic capacitance was estimated to be 6 fF. The low resistance benefits from the n-down epitaxial structure, as previously explained. The pad capacitance and inductance were estimated to be 8.5 fF and 42 pH, respectively. By optimizing the pad capacitance (19.5 fF) and inductance (65 pH), the 3-dB bandwidth was predicted to reach 220 GHz. More details are provided in Supplementary Note 6. Additionally, a 1 × N PD array was developed to enhance the output power. The 1 × 4 PD array exhibited a 45 GHz 3-dB bandwidth (see Supplementary Note 7). Using a heterodyne setup, we fixed the beat frequency while varying the optical power fed into the device to measure both the photocurrent and RF output power. Figure 2g illustrates that the 1 × 4 PD array achieved an RF output power of 0.2 dBm at 50 GHz, -1.7 dBm at 100 GHz, and −8.5 dBm at 150 GHz, all without requiring active cooling. This power handling capacity of the photodiodes satisfies the requirement of coherent optical communication and enables RF-amplifier-free signal reception. For the application of microwave/mmWave/THz photonics, these high-power and high-speed photodiodes enhance the system performance, including improved system gain, noise figure, and dynamic range[39], higher mmWave/THz signal power, and larger sensing and communication distance.

Figure 3a illustrates the fabricated optical coherent receiver, which consists of a 90° optical hybrid and a balanced PD array. The optical signal and the optical local oscillator (LO) are fed into an optical hybrid and subsequently detected by the balanced PD array (the pair of PD1-PD2 as the in-phase component and the pair of PD3-PD4 as the quadrature component). The generated I–Q electrical signals are processed via digital signal processing (DSP) to recover the original information. As a crucial component in the coherent receiver, the optical 90° hybrid (Fig. 3c) incorporates a paired-interference-based 2 × 4 MMI coupler and a general-interference-based 2 × 2 MMI coupler. The compact optical 90° hybrid features a 74 μm long 2 × 4 wedge-shaped MMI coupler and a 59 μm long 2 × 2 MMI coupler. The symmetry of the structure ensures that two adjacent outputs from the 2 × 4 MMI coupler maintain an in-phase relationship. A general-interference-based 2 × 2 MMI coupler was implemented to adjust the phase relationship of the pair of outputs by 90°. This configuration ensures four output signals with relative phase differences of 0°, 180°, 90°, and 270°. This eliminates the need for cross-waveguides required by conventional 4 × 4 MMI 90° hybrids, making it ideal for compact integrated coherent receiver chips[40]. Optical simulations of the optical 90° hybrid were conducted using the Lumerical finite difference time domain (FDTD) solver (Lumerical 2023 R1). Figure 3d, e illustrate the simulated optical power evolution of the optical 90° hybrid with two optical inputs. The optical power is evenly distributed to the four output ports for each of the optical inputs. The performance of the compact wedge-shaped optical 90° hybrid was experimentally characterized using a Mach–Zehnder interferometer (MZI) (see Supplementary Note 8). The device exhibited a phase deviation of less than ±5 degree (Fig. 3g), an excess loss of less than 0.5 dB (Fig. 3h), and a CMRR exceeding 20 dB (Fig. 3i) across the spectral range of 1530 nm to 1560 nm. The performance is comparable to commercial 90° optical hybrids[41].

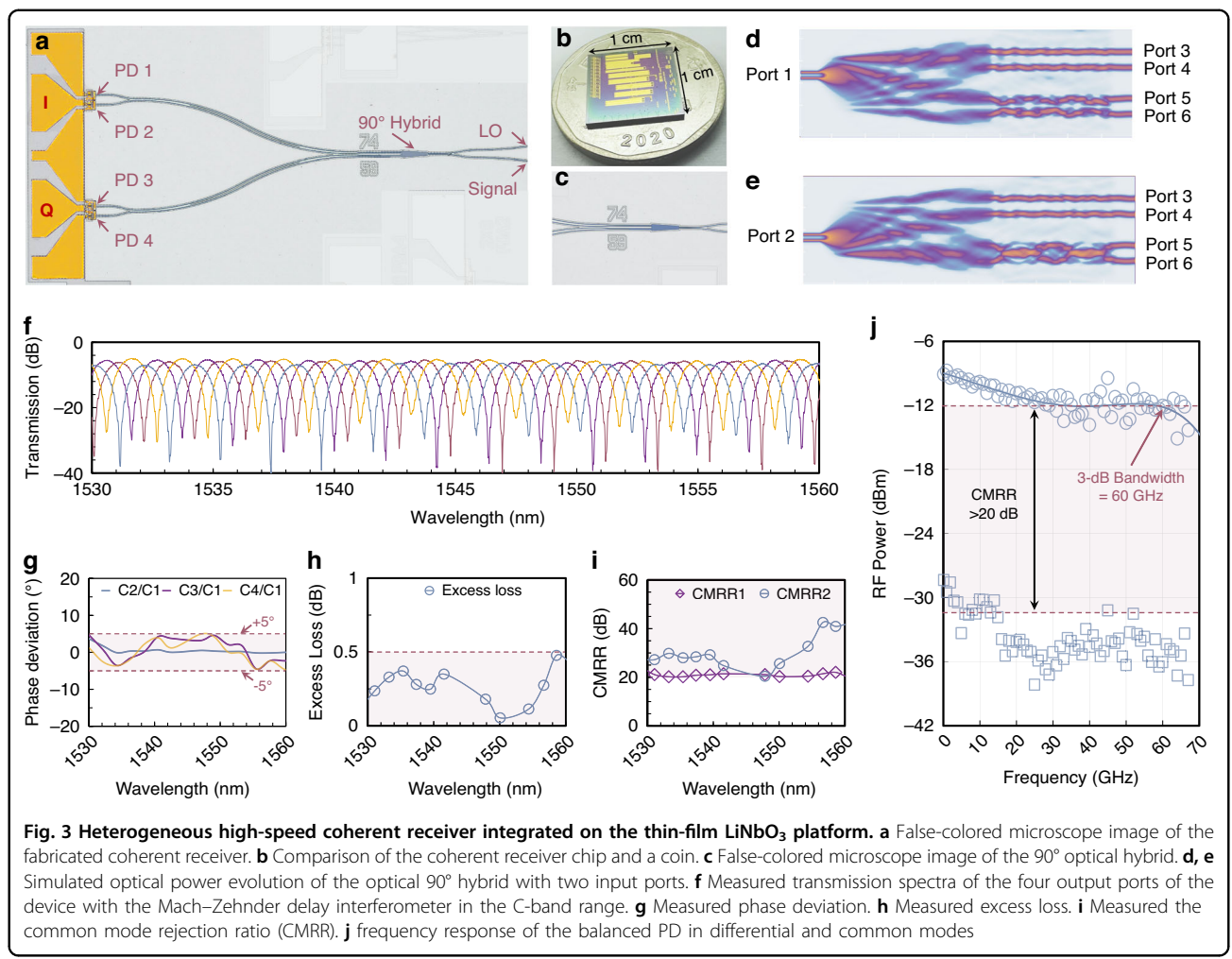

**Fig. 3 Heterogeneous high-speed coherent receiver integrated on the thin-film LiNbO₃ platform. a** False-colored microscope image of the fabricated coherent receiver. **b** Comparison of the coherent receiver chip and a coin. **c** False-colored microscope image of the 90° optical hybrid. **d, e** Simulated optical power evolution of the optical 90° hybrid with two input ports. **f** Measured transmission spectra of the four output ports of the device with the Mach–Zehnder delay interferometer in the C-band range. **g** Measured phase deviation. **h** Measured excess loss. **i** Measured the common mode rejection ratio (CMRR). **j** frequency response of the balanced PD in differential and common modes

An optical coherent receiver relies on a balanced PD with a large bandwidth and high CMRR to capture high-speed electrical signals with an optimal signal-to-noise ratio. The bandwidth and CMRR were characterized using a heterodyne setup with variable optical delay lines (see Supplementary Note 4). Figure 3j illustrates the frequency responses of a balanced PD in differential and common modes, revealing a 3-dB bandwidth of 60 GHz and a CMRR exceeding 20 dB. The high-performance

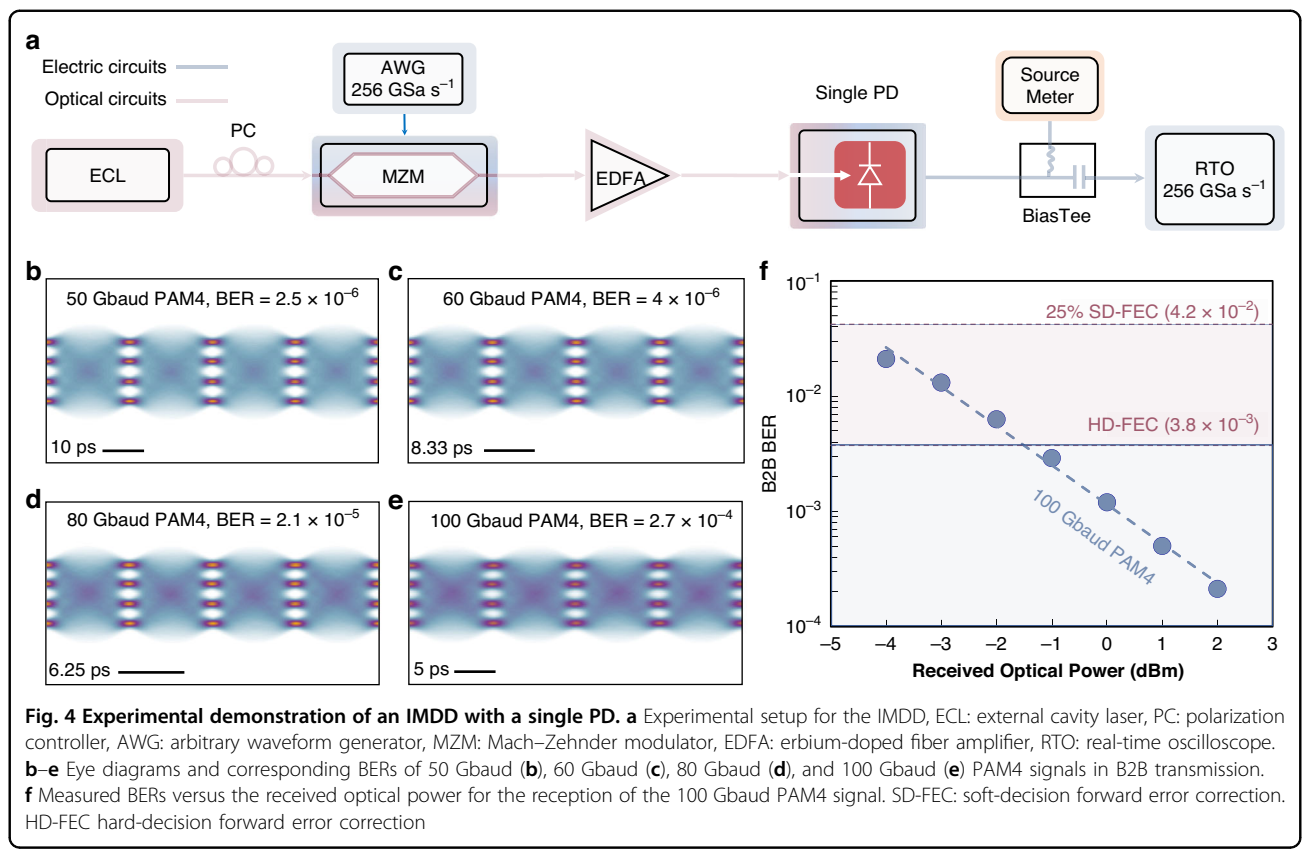

**Fig. 4 Experimental demonstration of an IMDD with a single PD. a** Experimental setup for the IMDD, ECL: external cavity laser, PC: polarization controller, AWG: arbitrary waveform generator, MZM: Mach–Zehnder modulator, EDFA: erbium-doped fiber amplifier, RTO: real-time oscilloscope. **b–e** Eye diagrams and corresponding BERs of 50 Gbaud (**b**), 60 Gbaud (**c**), 80 Gbaud (**d**), and 100 Gbaud (**e**) PAM4 signals in B2B transmission. **f** Measured BERs versus the received optical power for the reception of the 100 Gbaud PAM4 signal. SD-FEC: soft-decision forward error correction. HD-FEC hard-decision forward error correction

balanced PD lays the groundwork for realizing an ultrahigh-speed optical coherent receiver. The measured CMRR is generally sufficient to ensure robust signal recovery by suppressing common-mode intensity noise and maintaining high signal fidelity. Further improvement may be necessary for new applications such as quantum sensing and communication by minimizing fabrication-induced asymmetries in photodiode responsivities, optimizing electrode design with balanced impedance and phase matching, and implementing on-chip optical power and phase control[42].

**Intensity modulation direct detection (IMDD)**

To validate the performance of the fabricated devices, we applied a single PD in an IMDD system for the application of optical interconnect between intra-datacenter as shown in Fig. 1b. The experimental setup is depicted in Fig. 4a (see "Methods" section). The single PD successfully demodulated PAM4 signals with high baud rates at a 1550-nm wavelength. Figure 4b–e present the eye diagrams and corresponding bit error rates (BERs) for back-to-back (B2B) transmissions of 50 Gbaud, 60 Gbaud, 80 Gbaud, and 100 Gbaud PAM4 signals, respectively. All eye diagrams prove high-quality reception of the PAM4 signals, with clear delineation of the four-level symbols. Specifically, the BERs for the 50 Gbaud, 60 Gbaud, and 80 Gbaud PAM4 signals were below the KP-FEC limit of $2 \times 10^{-4}$, corresponding to data rates of 100 Gbit s$^{-1}$, 120 Gbit s$^{-1}$, and 160 Gbit s$^{-1}$, respectively. At a symbol rate of 100 Gbaud (200 Gbit s$^{-1}$), the device achieved high-performance reception with a BER of $2.7 \times 10^{-4}$, which was lower than the hard-decision forward error correction (HD-FEC) limit ($3.8 \times 10^{-3}$). Subsequently, the relationship between the BER and received optical power for the 100 Gbaud PAM4 signal was characterized by varying the incident optical power, as shown in Fig. 4f. When the received optical power reached −4 dBm, the BER was lower than the 25% soft-decision forward error correction (SD-FEC) limit of $4 \times 10^{-2}$, and when the received optical power was −1 dBm, the BER was lower than the HD-FEC limit.

**Coherent detection**

To verify the performance of the single-polarization coherent receiver for the application of optical interconnect between inter-datacenter as shown in Fig. 1b, we constructed the experimental setup depicted in Fig. 5a. The optical signal was modulated by the RF signal through an I–Q modulator with a bandwidth of 30 GHz. Following mixing in the optical 90° hybrid, the optical signal and LO were split into four outputs with relative phase differences of 0°, 180°, 90°, and 270°, which were coherently detected by a balanced PD array. The generated I–Q electrical signals were directly fed to a real-time oscilloscope (RTO) without the use of RF amplifiers. Details of the experimental setup are provided in the "Methods" section. We conducted coherent reception experiments at 1,550 nm with various symbol rates (80 Gbaud, 100 Gbaud, 128 Gbaud) and advanced modulation formats (QPSK, 16-QAM, 32-QAM, and 64-QAM). The received signals were processed by offline DSP (see "Methods" section) to calculate their BERs. Figure 5b summarizes the constellation diagrams and BERs measured for different symbol rates and modulation formats in B2B transmission. For a QPSK signal, error-free reception was achieved at a rate of 100 Gbaud (200 Gbit s$^{-1}$). For a QPSK signal with a symbol rate of 128 Gbaud (256 Gbit s$^{-1}$), the corresponding BER was $1.7 \times 10^{-5}$, which was below the KP-FEC limit. The measured BERs for the 80 Gbaud and 100 Gbaud 16-QAM signals were $1 \times 10^{-3}$ and $2.3 \times 10^{-3}$, respectively; both are below the HD-FEC limit, and the corresponding data rates were 320 Gbit s$^{-1}$ and 400 Gbit s$^{-1}$, respectively. Moreover, successful reception with BERs below the 25% SD-FEC limit was achieved for 128 Gbaud 16-QAM (BER = $3.4 \times 10^{-2}$), 80-Gbaud 32-QAM (BER = $1.1 \times 10^{-2}$), 100 Gbaud 32-QAM (BER = $2.6 \times 10^{-2}$), and 80 Gbaud 64-QAM signals (BER = $3.1 \times 10^{-2}$), with the corresponding highest data rate up to 512 Gbit s$^{-1}$. Furthermore, an experimental exploration of the reception of a 100 Gbaud 64-QAM signal, corresponding to a data rate of 600 Gbit s$^{-1}$, yielded a BER of $4.9 \times 10^{-2}$, lower than the 28% SD-FEC limit ($5 \times 10^{-2}$)[43].

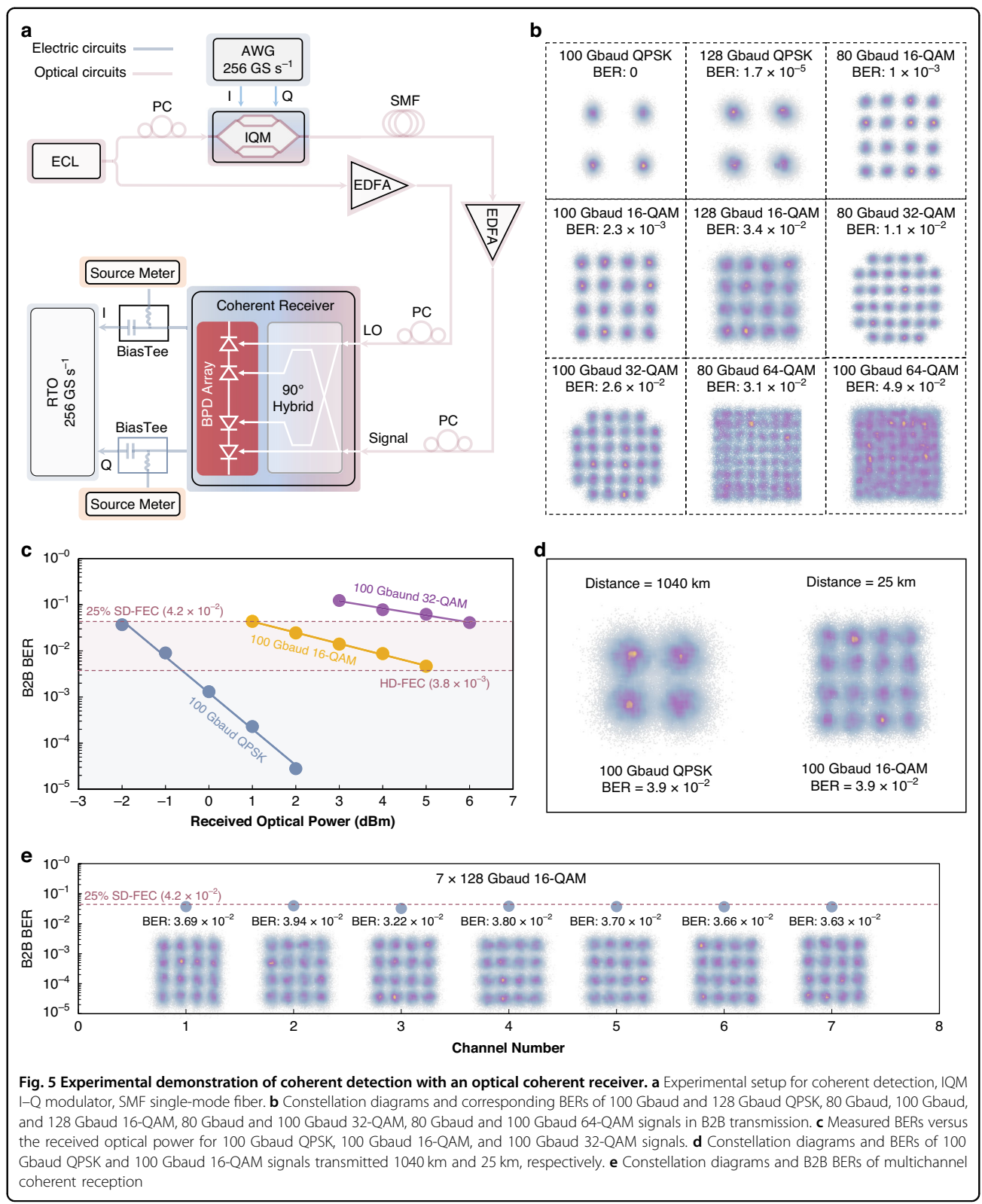

**Fig. 5 Experimental demonstration of coherent detection with an optical coherent receiver. a** Experimental setup for coherent detection, IQM I–Q modulator, SMF single-mode fiber. **b** Constellation diagrams and corresponding BERs of 100 Gbaud and 128 Gbaud QPSK, 80 Gbaud, 100 Gbaud, and 128 Gbaud 16-QAM, 80 Gbaud and 100 Gbaud 32-QAM, 80 Gbaud and 100 Gbaud 64-QAM signals in B2B transmission. **c** Measured BERs versus the received optical power for 100 Gbaud QPSK, 100 Gbaud 16-QAM, and 100 Gbaud 32-QAM signals. **d** Constellation diagrams and BERs of 100 Gbaud QPSK and 100 Gbaud 16-QAM signals transmitted 1040 km and 25 km, respectively. **e** Constellation diagrams and B2B BERs of multichannel coherent reception

As shown in Fig. 5c, the relationship between the BER and received optical power was determined for 100 Gbaud QPSK, 100 Gbaud 16-QAM, and 100 Gbaud 32-QAM signals. For 100 Gbaud QPSK signals, when the received optical power reached −2 dBm, 0 dBm, and 2 dBm, the measured BERs were below the 25% SD-FEC limit, HD-FEC limit, and KP-FEC limit, respectively. For the 100 Gbaud 16-QAM and 100 Gbaud 32-QAM signals, the received BERs were below the 25% SD-FEC threshold when the incident power of the PD reached 1 dBm and 6 dBm, respectively.

Furthermore, we conducted long-distance transmission communication experiments using a coherent receiver chip. The experimental results are shown in Fig. 5d. After 1040 km single-mode fiber transmission, the BER of the received 100 Gbaud QPSK signal was $3.9 \times 10^{-2}$. For the 100 Gbaud 16-QAM signal, the BER was $3.9 \times 10^{-2}$ after 25 km single-mode fiber transmission. In the long-distance communication experiment, optical amplification (16 dB gain) was performed every 80 km fiber transmission (~16 dB loss). Finally, a seven-channel single-polarization I–Q coherent receiving chip was employed to perform a multichannel communication demonstration in B2B transmission. With the 25% SD-FEC limit, 128 Gbaud 16-QAM signals were successfully received for all the channels, pushing the total communication capacity of a single chip to 3.584 Tbit s$^{-1}$, as shown in Fig. 5e.

### Discussion

As presented above, a wafer-level heterogeneous InP-LiNbO$_3$ integration platform has been developed, enabling large-scale manufacturing with high process yield and multifunctionality integration. The single PD achieved a record-high bandwidth of 140 GHz on the thin-film LiNbO$_3$ heterogeneous integration platform, which was attributed to the minimized parasitic capacitance and series resistance. Through detailed engineering of the resonant effect within the PD's equivalent circuit, we project that the 3-dB bandwidth could exceed 200 GHz. This advancement opens the door to future short-range intensity-modulated direct detection (IMDD) optical interconnects with symbol rates surpassing 300 Gbaud. In addition to the high bandwidth capability, the carefully designed epitaxial structure of the PDs enhances output RF power, facilitating RF-amplifier-free direct or coherent reception of high-speed signals while enhancing the microwave/mmWave/THz photonics system performance. The integration of mmWave/THz modulators and photodiodes enables optical interconnects between data centers, central stations, and base stations, as well as facilitating mmWave/THz signal emission and reception, as illustrated in Fig. 1b.

The thin-film LiNbO$_3$ optical coherent receiver, enabled by a high-performance 90° optical hybrid and high-speed balanced PD array, demonstrates exceptional performance in receiving ultrahigh-speed signals with advanced modulation formats. To the best of our knowledge, this is the first experimental demonstration of an ultrahigh-speed coherent receiver on a thin-film LiNbO$_3$ platform. Table 1 presents a comparative overview of the state-of-the-art on-chip integrated optical coherent receivers reported so far[44–49], including monolithic and heterogeneous integration platforms based on materials such as InP/InGaAs, SiGe, graphene, and LiNbO$_3$. Compared with other integration platforms, our heterogeneous integrated, coherent receiver achieves a 75% improvement of 3-dB bandwidth through the careful design of a modified uni-traveling-carrier photodiode epitaxy structure with an n-mesa-down configuration.

In our work, the single-polarization coherent receiver chip demonstrates powerful capability in demodulating ultrahigh-speed signals with advanced modulation formats. It achieved a maximum data rate of 600 Gbit s$^{-1}\lambda^{-1}$Pol$^{-1}$ (100 Gbaud 64-QAM) in B2B transmission. Multichannel coherent reception of a 7 × 128 Gbaud 16-QAM signal further underscores its potential for ultra-large capacity short-range optical coherent interconnects, delivering a total data capacity of 3.584 Tbit s$^{-1}$. Moreover, the coherent receiver demonstrates application prospects in long-distance optical communication systems, successfully receiving 100-Gbaud QPSK signals over 1040 km and 100-Gbaud 16-QAM signals over 25 km. It is noteworthy that the reception capacity is currently constrained by the bandwidth of the available I–Q modulator and AWG. Future enhancements could leverage higher bandwidth I–Q modulators and more advanced AWG to achieve even larger capacity. Lastly but most importantly, the coherent receiver achieves these record-large capacities without the need for RF amplification, resulting in ultra-low energy consumption. Our developed heterogeneous integrated coherent receiver consists of two on-chip balanced photodiodes. Only two-channel high-speed electronic components per polarization are required, significantly reducing the power consumption and complexity of the electronic circuitry. Specifically, energy consumption as low as 9.6 fJ bit$^{-1}$ (for 200 Gbit s$^{-1}$) and 13.5 fJ bit$^{-1}$ (for 400 Gbit s$^{-1}$) highlights its efficiency in high-speed data transmission scenarios (see Supplementary Note 9).

Table 1  Literature overview of the state-of-the-art integrated optical coherent receivers

| Year/Ref | 2020[44] | 2020[45] | 2021[46] | 2022[47] | 2023[48] | 2024[49] | This work |
|---|---|---|---|---|---|---|---|
| Material | InP | SiGe | Graphene/SOI | SiGe | SiGe | InP | InP/thin-film LiNbO$_3$ |
| Integration Method | Monolithic | Monolithic | Heterogeneous | Monolithic | Monolithic | Monolithic | Heterogeneous |
| Optical Band | C-band | C-band | C-band | C-band O-band | O-band | C-band | C-band |
| Polarization No. | 2 | 2 | 1 | 1 | 2 | 1 | 1 |
| Channel No. | 2 | 1 | 1 | 1 | 1 | 1 | 1 |
| 3-dB BW (GHz) | 50 (Single PD) | 52 (Single PD) | > 67 (Single PD) | 33 (Single PD) | 42 (Single PD) | 80 (Single PD) | 140 (Single PD) 60 (Balanced PD) |
| Symbol Rate (GBaud) | 90 | 96 | 100 | 64 | 66 | 128 | 128 |
| Modulation Format | 64 QAM | 16 QAM | QPSK 16 QAM | QPSK | 16 QAM | 16 QAM | 16 QAM 32 QAM 64 QAM QPSK 16 QAM |
| Bit Rate (Gbit s$^{-1}$λ$^{-1}$Pol$^{-1}$) | 400 | 384 | 200 240 | 128 | 264 | 512 | 512 500 600 200 400 |
| Distance (km) | B2B | B2B | B2B B2B | B2B | B2B | B2B | B2B B2B B2B 1040 25 |

In conclusion, our proposed thin-film LiNbO$_3$ heterogeneous integrated PD achieves a record-high bandwidth of 140 GHz. The optical coherent receiver features a balanced detection bandwidth of 60 GHz, a CMRR of > 20 dB, and an energy consumption of 9.6 fJ bit$^{-1}$. This capability supports data reception of up to 600 Gbit s$^{-1}$ per polarization channel and achieves a total capacity of 3.584 Tbit s$^{-1}$ with seven channels. Compared to existing integrated optical coherent receivers, our heterogeneous integrated coherent receiver sets new benchmarks with its superior bandwidth, ultra-low energy consumption, and ultra-large capacity. Furthermore, our InP-LiNbO$_3$ heterogeneous integration platform has the capability to integrate lasers, modulators, and receivers on a single LN chip by converging wafer bonding and selective regrowth techniques[50]. The integrated transceiver chip would significantly reduce the package assembly complexity and increase the integration density of the entire chip. Beyond the high-performance transceiver for hyperscale datacenter, the multi-function InP-LiNbO$_3$ wafer-level heterogeneous integration platform has the potential to enable microwave/mmWave/THz photonics signal generation and manipulation chips and full-photonics THz communication and sensing integrated chips.

## Materials and methods
### Device fabrication
All patterning and alignment throughout the manufacturing process were performed on thin-film LiNbO$_3$ wafers and bonded wafers using an i-line stepper lithography system. Thin-film LiNbO$_3$ waveguides and passive devices were precisely crafted using inductively coupled plasma (ICP) with argon to etch 300 nm of LiNbO$_3$, leaving a 300 nm slab layer intact. Prior to bonding, the LiNbO$_3$ surface underwent chemical mechanical polish to ensure a smooth, uniform surface and to minimize the impact of nanoscale waveguide features, which was critical for subsequent wafer bonding processes. The 2-inch InP/InGaAs photodiode wafer was bonded onto the 4-inch LNOI wafer. After wafer bonding, InP substrate removal involved grinding and selective wet etching with HCl/H$_2$O, carefully stopping at the InGaAs p-contact layer. An electron-beam evaporator deposited a p-metal stack (Ti/Pt/Au/Ti) to establish ohmic contacts. The p-mesa was etched to the n-contact layer by using a Cl$_2$-based recipe with an ICP etcher. Subsequently, an n-contact metal (AuGe/Ni/Au) was deposited via electron-beam evaporation and a lift-off process. The same Cl$_2$-based dry etching process was employed to define the n-mesa, with final exposure of the LiNbO$_3$ surface achieved through wet etching with HCl/H$_2$O. The LiNbO$_3$ surface was further treated with diluted hydrofluoric acid to eliminate potential contaminants and surface damage. A 600 nm SiO$_2$ layer was deposited across the entire wafer surface using plasma-enhanced chemical vapor deposition (PECVD) as a passivation layer. This layer serves multiple purposes: reducing dark current, safeguarding lithium niobate waveguides and passive devices, and mitigating the impact of impurity particles. Etching of the SiO$_2$ layer above the p- and n-metal stacks was performed using an ICP etching system with CF$_4$ gas, followed by the formation of metal electrodes through electroplating and lift-off processes. After completing PD fabrication, the SiO$_2$ passivation layer in the modulator region was dry-etched first and then wet-etched to expose the LN waveguide. The 1-μm-thick metal (Ti/Au) was deposited on the LN to form a transmission line. A 900-nm SiO$_2$ cladding layer was deposited on the whole wafer to protect the waveguide and tune the optical group index and then was opened by dry-etch in the RF pad area. Finally, the wafer was diced into small chips with size of 1 × 1 cm. Due to the use of edge coupling, the edges of the chips were side-polished.

### Device characterization

The experimental setup for measuring the bandwidth of a single PD is detailed in Supplementary Note 4 Fig. S6a. Two external cavity lasers (Keysight 81940A) were combined using a 3 dB coupler to generate an optical beat signal with 100% modulation depth. The beat signal frequency was adjusted by controlling the wavelength difference between the lasers. Optical power was amplified using a high-power EDFA and then coupled into a LiNbO$_3$ waveguide via a lensed fiber with a 2.5 μm spot diameter. The optical signal was detected by a single PD. The frequency response of the devices was characterized using ground-signal-ground (GSG) probes covering the DC–67 GHz, 75 GHz–110 GHz, 90 GHz–140 GHz, and 110 GHz–170 GHz frequency bands. Supplementary Note 4 Fig. S6b, illustrates the setup for measuring the frequency response of the balanced PD. Optical beat signals were split into two paths, each incorporating a variable optical delay line to adjust the phase difference between them. Common-mode (even multiples of the π phase difference) and differential-mode (odd multiples of the π phase difference) were obtained by tuning the phase difference. Variable attenuators were employed to compensate for power imbalances resulting from losses in the optical paths. Optical signals from each path were coupled to the balanced PD via lensed fibers. A DC–67 GHz RF probe was used to measure the bandwidth of the balanced PD. For all bandwidth measurements, the bias voltage supplied by a source meter (Keithley 2400) through a bias tee (SHF BT65R, 65 GHz) was applied to the PDs in the DC–67 GHz frequency range. In the 75–110 GHz, 90–140 GHz, and 110–170 GHz frequency ranges, internal bias-tees integrated into the waveguide RF probes provided the bias voltage to the single PDs. The output photocurrent and RF power were measured using a source meter and an RF power meter (Ceyear 2438CA). Calibration accounted for additional losses from RF probes, bias tees, and RF cables to ensure accurate bandwidth and output power measurements.

For the characterization of the intensity modulator, a vector network analyzer (VNA Ceyear 3672) was used to measure the S parameters. The characteristic impedance and RF effective index can be extracted through S parameters. The EO response of the intensity modulator was measured by an optical measurement setup. A 1550-nm laser was used to input light to the chip via a lensed fiber with a 2.5 μm spot diameter. The output light of the modulator was received by a high-speed photodetector (Finisar XPDV2120RA). The VNA was used to measure the total RF response (DC-50GHz) of the optical link. The EO response of the intensity modulator can be obtained by de-embedding OE response of the photodiode. The half-wave voltage of the modulator was characterized by the triangular wave method.

### High-speed data reception for IMDD and coherent detection

In the IMDD experiment, a PAM4 signal with a length of $2^{20}$ data cycles was generated from an AWG (Keysight M9505A, 256 GSa s$^{-1}$) and modulated on the signal light using an I–Q modulator with a 30 GHz bandwidth (EOspace 50 Gbaud I–Q modulator). The polarization state of the light in the I–Q modulator was controlled by a polarization controller to maximize the output power of the I–Q modulator. Subsequently, the modulated light was amplified by an EDFA and coupled into the single PD via a lensed fiber.

For the coherent reception experiment, RF signals with amplitude and phase information from the I and Q channels were modulated to the signal light. Both the signal light and LO light were amplified by the EDFAs and coupled into the thin-film LiNbO$_3$ coherent receiver chip via lensed fibers with a spot size of 2.5 μm. The power of the LO light was adjusted to 8 dB higher than that of the signal light. Following photonic mixing within the on-chip optical 90° hybrid, the lights were split and fed into the balanced PD array, generating I and Q electrical signals. Bias voltages for the balanced PD array were applied using a customized ground-signal-ground-signal-ground (GSGSG) RF probe covering a frequency range from DC to 67 GHz. The electrical signals from the balanced PD array were directly sent to a real-time oscilloscope (Keysight UXR0594AP, 256 GSa s$^{-1}$) without additional RF amplification. The eye diagrams, constellation diagrams, and BERs were calculated by an offline DSP.

### Digital signal processing

For PAM4 signal processing in DSP, the collected samples were first resampled by a factor of 4. The Gardner algorithm was subsequently used to recover the clock, followed by decision-directed minimum mean square equalization to recover the signal. Finally, BERs were calculated through symbol-to-bit mapping. For coherent signal processing in DSP, I–Q modulated signals were first processed using the Schmidt orthogonalization algorithm to compensate for the non-orthogonality between the two signal paths caused by the modulator and the 90° optical hybrid coupler, thereby mitigating the associated impairments. Then the signals were resampled at two samples per symbol. Clock recovery and constant modulus algorithm (CMA) equalization were applied. Carrier phase estimation (CPE) was subsequently performed to remove phase noise from the signal. The decision-directed least mean square (DDLMS) algorithm was used for equalization to further reduce inter-symbol interference. Finally, BERs were computed by aligning the received signal with a reference signal.

**Acknowledgements**
This work was supported by the National Key Research and Development Program (Grant No. 2022YFB2803800) and the National Natural Science Foundation of China (Grant No. U23A20376, 62431024). We sincerely thank Prof. Zhongming Zeng (School of Nano Technology and Nano Bionics, University of Science and Technology of China) for assistance with chip fabrication, Prof. Haijun He for assistance with the optical communication test, and Tengmu Chen and Shiyong Peng for assistance with device characterization.

**Author contributions**
X.X. and C.W. contributed equally to this work. X.X. proposed the original concept and designed the PD epi layers and the whole fabrication process. C.W. partially designed the fabrication process and fabricated the wafer by hand. Y.C. designed and measured the MMI couplers and 90° optical hybrid. X.H. designed the equalization algorithm for high-speed coherent signal reception. C.W., X.H., C.W., and J.S. performed the IMDD and coherent communication measurements. C.W. and X.X. prepared the manuscript. L.J., J.Y., X.Z., W.P., and L.Y. revised the manuscript and provided insightful comments. X.X. and L.Y. supervised the research.